\documentclass[preprint,times]{aastex}
\usepackage{natbib}
\usepackage[usenames]{color}

\begin{document}

\title{Comet 17P/Holmes in  Outburst: The Near Infrared Spectrum}
\author{Bin Yang, David Jewitt and Schelte J. Bus
\affil{Institute for Astronomy, University of Hawaii, Honolulu, HI
96822} \email{yangbin@ifa.hawaii.edu, jewitt@ifa.hawaii.edu}}
\maketitle

\begin{quotation}
\begin{center}
{\bf Abstract }\\
\end{center}
Jupiter family comet 17P/Holmes underwent a remarkable outburst on UT 2007 Oct.\ 24,
in which the integrated brightness abruptly increased by about a factor of a million.
 We obtained near infrared  (0.8 - 4.2$\mu$m) spectra of 17P/Holmes on UT 2007 Oct.\ 27, 28 and 31, using the 
3.0-m NASA Infrared Telescope Facility (IRTF) atop Mauna Kea.  Two broad absorption bands 
were found in the reflectance spectra with centers (at 2$\mu$m 
and 3$\mu$m, respectively) and overall shapes consistent with the presence of water ice grains in the coma. 
Synthetic mixing models of these bands suggest an origin in cold ice grains of micron size.  Curiously, though, the expected 1.5$\mu$m band of water ice 
was not detected in our data, an observation for which we have no explanation.  Simultaneously, 
excess thermal emission in the spectra at wavelengths beyond 3.2$\mu$m has a color temperature of  
360 $\pm$ 40 K (corresponding to a superheat factor of $\sim$ 2.0 $\pm$ 0.2 at 2.45 AU). This is too 
hot for these grains to be icy.  The detection of both water ice spectral features and short-wavelength thermal emission suggests that 
the coma of 17P/Holmes has two components (hot, refractory dust and 
cold ice grains) which are not in thermal contact. A similarity to grains ejected into the coma of 
9P/Tempel~1 by the Deep Impact spacecraft is noted. \end{quotation}

\section{Introduction}
As a Jupiter-family comet, 17P/Holmes is unremarkable in terms of either its dynamical properties or its chemical composition \citep{whipple:1984}. However, on UT 2007 Oct.\ 24, its brightness increased by about a million times in less than a day, with its apparent magnitude rising from $\sim$17 to $\sim$1.5  \citep{buzzi:2007}. In comparison, NASA's Deep Impact into the nucleus of comet 9P/Tempel~1 in 2005 changed the brightness of the comet by a comparatively modest two magnitudes \citep{fernandez:2007}.  The spectacular brightening of 17P echoed an earlier outburst on November 6 1892, which led to the discovery of the comet by the British amateur astronomer Edwin Holmes in London.  The two outbursts, 115 years apart, are strikingly similar in the sense that both occured 5 months after the comet had passed its perihelion and in both cases the cometary coma increased dramatically in angular size and brightness. However, whereas the comet experienced a second outburst in January 1893, two months after the initial event,  no such second outburst followed the October 2007 event. 

Over the years, various models and hypotheses have been advanced to explain cometary outburst phenomena \citep{2007AN....328..126G}.  Some outbursts are associated with cometary splitting, in which the primary nucleus ejects discrete fragments, perhaps in response to rotational or other instabilities.  Unfortunately, there is no strong evidence concerning the rotational period of 17P \citep{2006MNRAS.373.1590S} and nothing to suggest that it is close to rotational break-up.  Some comets split due to strong tidal forces from nearby massive objects, usually the Sun or Jupiter \citep{Boehnhardt:2004}. However, 17P/Holmes was far from Jupiter and the Sun at the time of the outbursts, giving no reason to suspect a tidal trigger. \citet{whipple:1984} described a scenario whereby the impact of an unseen satellite could have triggered the 1892 outburst. However, the repetitive nature of the outbursts in this comet renders this explanation implausible.  A conjectural scenario is that a large fraction of 17P/Holmes is amorphous water ice in which are trapped considerable amounts of volatile molecules. When close to the Sun, elevated internal temperatures might have precipitated an irreversible phase transition in water ice from amorphous to crystalline.  This (exothermic) transition process would release the trapped volatiles, driving up the gas pressure and eventually  triggering the observed outburst (e.g. Prialnik et al.\ 2004). 

In this paper we study the material ejected from 17P via near infrared spectroscopy (0.8 - 4.2$\mu$m) obtained shortly after the start of the outburst. 

\section{Observations}
Observations were obtained using the NASA Infrared Telescope Facility (IRTF) 3-m
telescope atop Mauna Kea, Hawaii. We observed 17P/Holmes on UT 2007 Oct.\ 27, 28 and 31, using SpeX \citep{rayner:2003}, a medium-resolution 0.8-5.5$\mu$m spectrograph. SpeX is equipped with a Raytheon 1024 $\times$1024 InSb array having a spatial scale of 0.$\!\!^{\prime\prime}$15 pixel$^{-1}$.  Two cross-dispersed modes,  known as SXD (0.8 - 2.4$\mu$m) and  LXD (1.9 - 4.2$\mu$m), were used to cover an overall wavelength range from 0.8$\mu$m to 4.2$\mu$m for all our IRTF observations. The spectral resolution afforded by SpeX is dependent, in part, on the width of the spectrograph slit.  To achieve high resolution,  we used a 0.$\!\!^{\prime\prime}$3$\times$ 15$''$  slit that provided an average spectral resolving power of $\lambda$/$\Delta$$\lambda$ = 2300. The slit was projected North-South on the sky. Differential refraction is minor because 17P/Holmes is an extended source. To avoid potential contamination from the coma, we used a large nod distance of 10 arcminutes to measure the sky background.   

\begin{figure}
\begin{center}
\includegraphics[width=4.5in,angle=90]{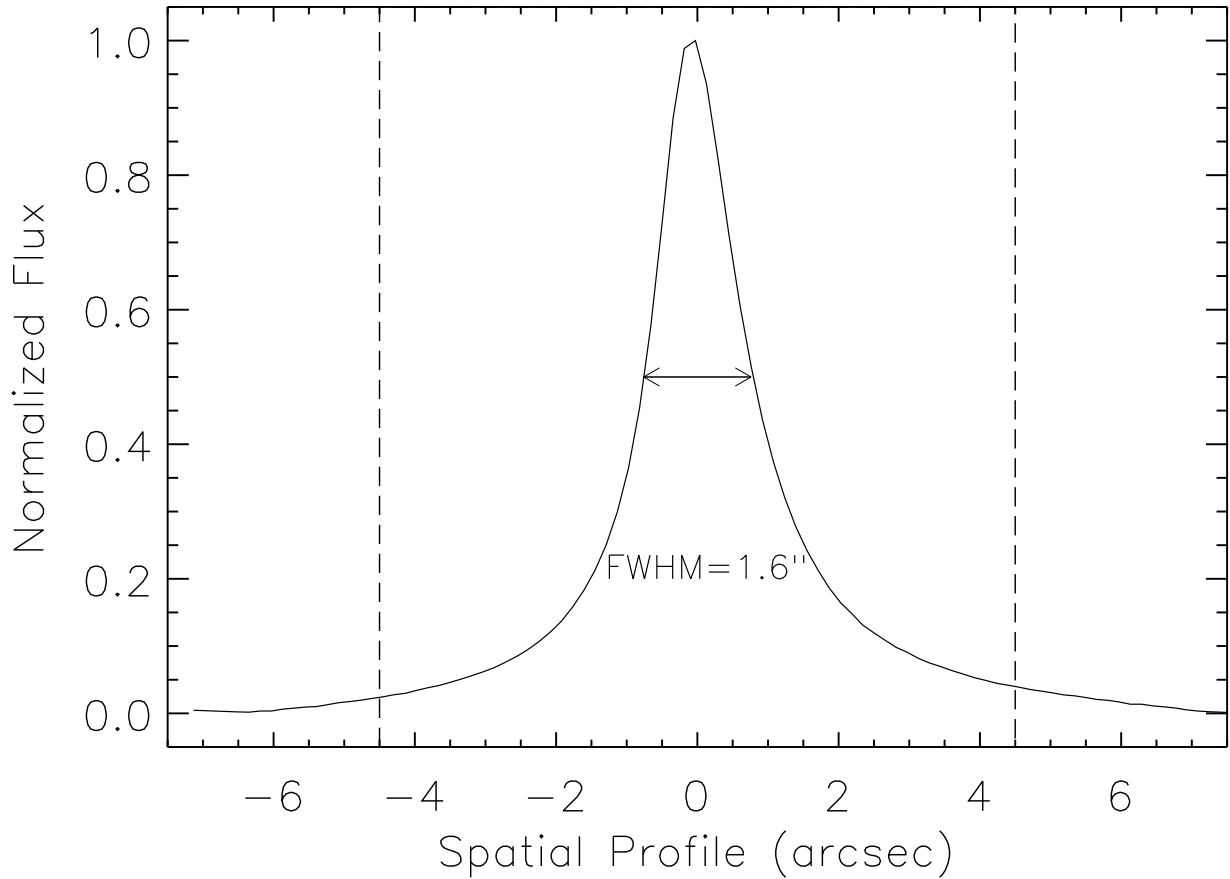}
\vspace{0.3in}
\caption{The solid line shows the spatial profile of the comet along the slit. The two dashed lines show the regio, of width of 9.0$''$, from which the final spectrum was extracted. Although the comet was surround by a sizable coma at the time of the observation, the surface brightness profile of 17P was centrally condensed, with a roughly Gaussian profile having a FWHM of 1.6$''$.}
\label{plotone}
\end{center}
\end{figure}

The SpeX data were reduced using the reduction pipeline SpeXtool \citep{Cushing:2004}. Individual flats were scaled to a
common median flux level, then combined and normalized to generate a master flat field frame. Bad pixels were identified as outliers using flat field frames as well as identified from a stored bad pixel mask. Pixels with poor response were replaced by interpolation over neighboring normal pixels. For standard stars, the first-order sky removal was achieved via subtracting image pairs, 7.5 arcsec.\ apart, with the object dithered along the slit.  For the comet, the first-order sky was removed by subtracting each sky (so-called B-beam) image, 10 arcminute East of the comet, from the corresponding object  (A-beam) image. The pixel-to-pixel sensitivity variations were calibrated by dividing the master flat field frame by the object frames. The wavelength calibration was accomplished using argon lines for the SXD mode, while a combination of argon lines and several sky emission lines were used for the LXD mode.  Finally, one-dimensional spectra of the comet and standard stars were extracted from the two-dimensional sky-subtracted images. The SpeXtool pipeline provides two extraction modes that are designed for point sources and extended sources respectively. The difference between these two modes is that the extended-source mode (ESM) sums up all the pixel counts uniformly within the extraction aperture, whereas the point-source mode (PSM) specifies the aperture center first and then carries out a  profile extraction, weighted by the spatial profile of the object along the slit. Although 17P/Holmes was, strictly speaking, an extended source, the spatial profile of this comet is in fact highly centrally-condensed and appeared roughly Gaussian (Figure~\ref{plotone}). Therefore, we performed test extractions using both ESM and PSM. We found that the two modes produced similar results, but the PSM provided smoother spectra with slightly higher signal-to-noise ratios (SNR). Therefore, we adopted the point source extraction mode using an average extraction aperture box width of 9.0$''$, to maximize the SNR.  

In order to remove telluric absorption features we obtained spectra of a ``telluric standard star''  close in both time
and sky position to the comet. We chose the A-type star BS 1261 \citep{1997A&A...323L..49P} as the telluric-calibration standard star, which also served as a reference to approximately flux calibrate the comet spectra. We chose an A-type star, because such early-type stars generally have few metal lines and can be well-modeled by a blackbody in the near infrared (e.g. Vacca et al. 2003).  Once the target spectra had been calibrated for telluric absorptions, the different spectral orders from SpeX were merged to construct a single, continuous spectrum covering the wavelength range from 0.8 to 4.2$\mu$m. However, the spectra at the longer wavelength end suffered from the rapid decrease in the sensitivity of the spectrograph and appeared noisy. Therefore, we discarded the low quality data in the wavelength region at 4.0$\mu$m to 4.2$\mu$m. Since no absolute flux-calibrated solar spectrum is available at wavelengths longer than 2.5$\mu$m, we used a spectrum of the solar analog HD 76151 from the IRTF spectral library (Rayner et al., 2008)\footnotemark\ as a substitute for the solar spectrum. As a check, we carefully examined the discrepancies between the absolute flux-calibrated solar spectrum (0.8 - 2.5$\mu$m) from \citet{Colina:1996} and the spectrum of HD 76151. We found that the difference between these two is less than 2\% over the 0.8$\mu$m to 2.5$\mu$m wavelength range, which is negligible for our present purposes.
\footnotetext{http://irtfweb.ifa.hawaii.edu/$\sim$spex/WebLibrary/index.html}

A journal of observations is provided  in Table 1.

\section{Analysis}
\subsection{Thermal Emission}
\begin{figure}
\begin{center}
\includegraphics[width=5.5in]{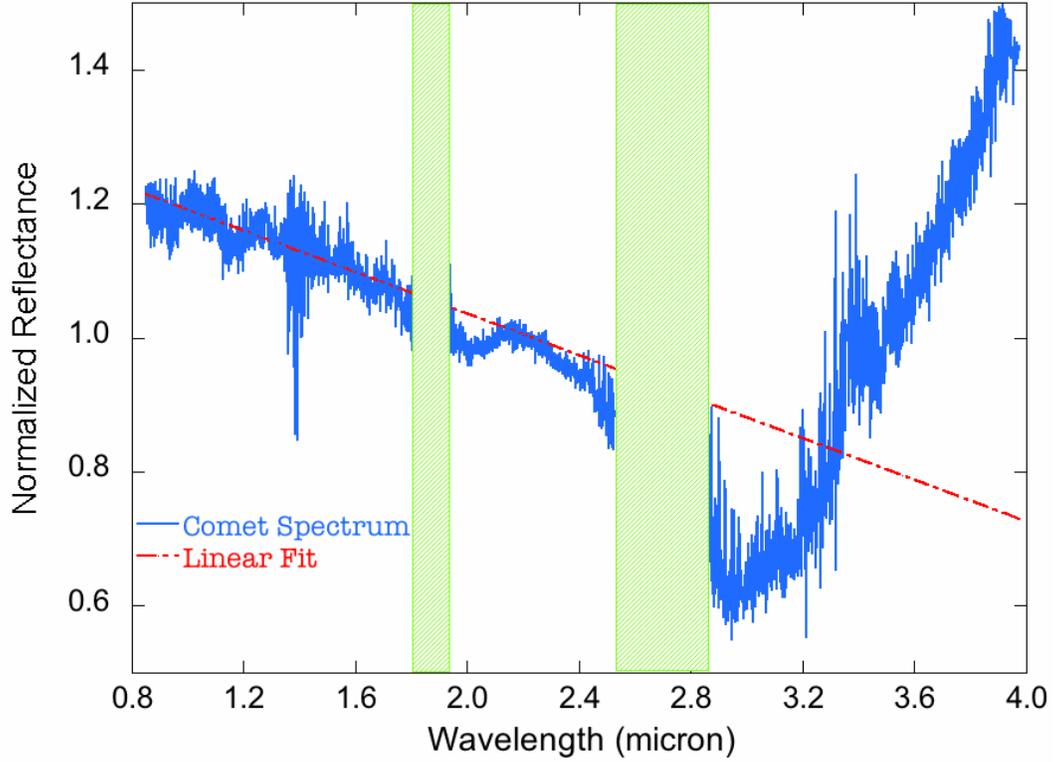}
\vspace{0.2in}
\caption{The reflectance spectrum of 17P/Holmes from 0.8- to 4.0-$\mu$m, taken on UT 2007 Oct.\ 27 and normalized at $\lambda = 2.2\mu m$.  At shorter wavelengths ($\lambda  <  $ 3.0$\mu$m), the most significant spectral feature of Holmes is the negatively sloped continuum. A linear fit to the continuum is shown as the red dashed line.  The rising spectrum at wavelengths ($\lambda > $ 3.2$\mu$m) is attributed to thermal emission. The two green shaded rectangles block out the regions that are heavily contaminated by the telluric absorptions.}
\label{plottwo}
\end{center}
\end{figure}

We computed reflectance spectra of the comet via dividing the flux calibrated spectra of 17P by the library spectrum of the solar substitute HD 76151 (Rayner et al., 2008). The reflectance spectra of 17P/Holmes rise sharply beyond 3.2$\mu$m (see Figure~\ref{plottwo}), which we interpret as the result of thermal emission from hot dust in the coma. Beyond 3.2$\mu$m, the spectrum is the sum of two parts, one from scattered sunlight and the other from thermal emission. The scattered and thermal components were modeled separately from each other. Firstly,  we assumed that the spectrum at  $\leqslant$ 2.5$\mu$m consists of pure scattered light and the scattering flux was estimated using an inverse-square law described in \citep{russell:1916}
\begin{equation}
\label{inverse}
f_{scat}(\lambda)=\frac{f_{\bigodot}(\lambda)p(\lambda)\phi(\alpha)C}{\pi r^2 \Delta^2}
\end{equation}

\noindent where $f_{scat}(\lambda)$ (W m$^{-2}$\AA$^{-1}$) is the flux density of scattered solar radiation and $f_{\bigodot}(\lambda)$ (W m$^{-2}$\AA$^{-1}$) is the flux density of the solar radiation at 1AU, $p(\lambda)$ is the geometric albedo of the grains in the comet and $\phi(\alpha)$ is the scattering phase function. Previous studies show that the cometary phase function can be adequately represented by: \begin{equation} \phi(\alpha) = 10 ^{-0.4 \beta \alpha}
\end{equation} where $\alpha$ (degree) is the phase angle and $\beta$ (mag deg$^{-1}$) is the linear phase coefficient \citep{Jewitt:1988}.  When the phase angle is small, $\alpha \leqslant $ 30$^\circ$, the measured $\beta$ $\approx$ 0.03 \citep{Meech:1987}.  In Equation (\ref{inverse}), $r$ (AU) and $\Delta$ (m) are the heliocentric and geocentric distances, respectively, and $C$ (m$^2$) is a scale factor. Physically,  $C$ is the total cross section of the cometary grains within the projected slit profile. Due to the uncertainty of grain sizes and albedo information, we treated $C$ as a free fitting parameter. 

\begin{figure}[h!]
 \begin{center}
 \includegraphics[width=6.5in]{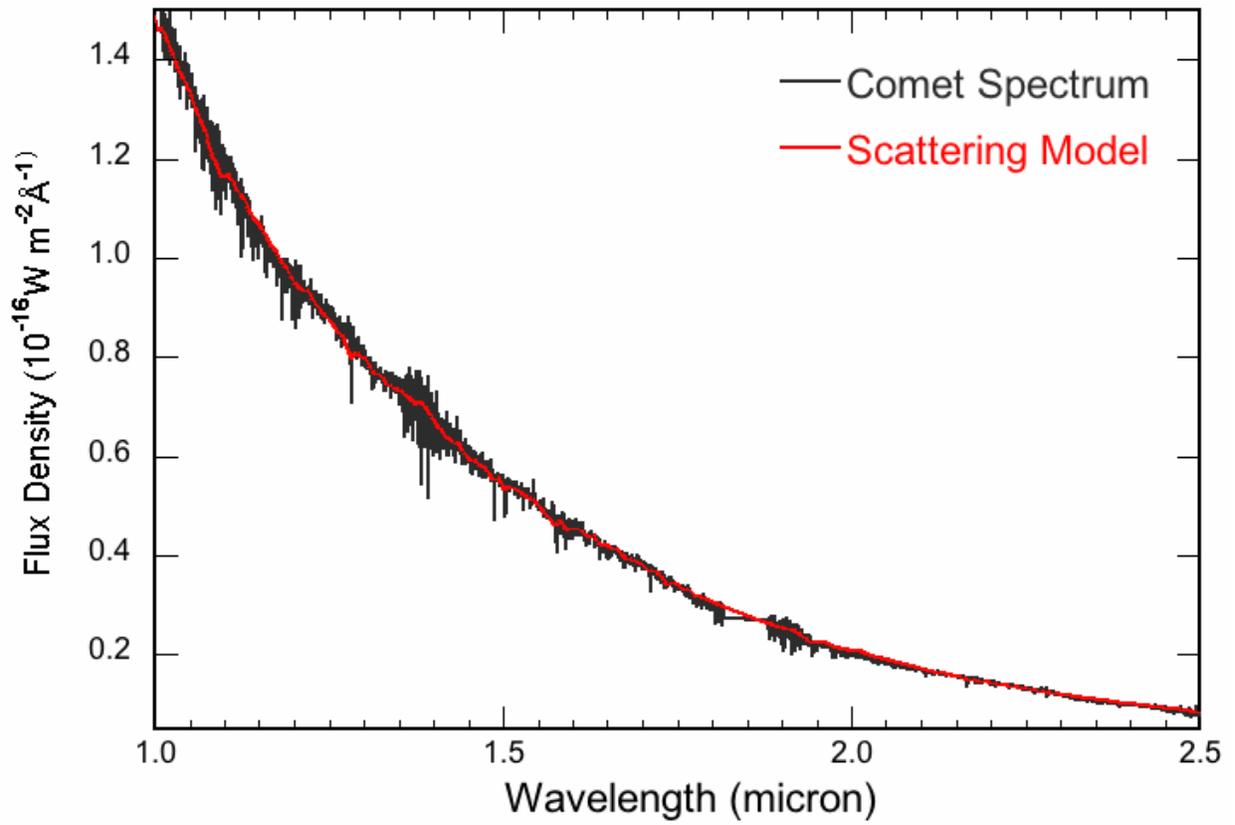}
\caption{The black line shows the comet spectrum, taken on UT 2007 Oct.\ 27. The red line represents the modeled scattering spectrum, which matches well with the observed data in the wavelength range 1.0 - 2.5$\mu$m. }
\label{plotthree}
\end{center}
\end{figure}

\begin{figure}[h!]
 \begin{center}
\includegraphics[width=4.5in,angle=90]{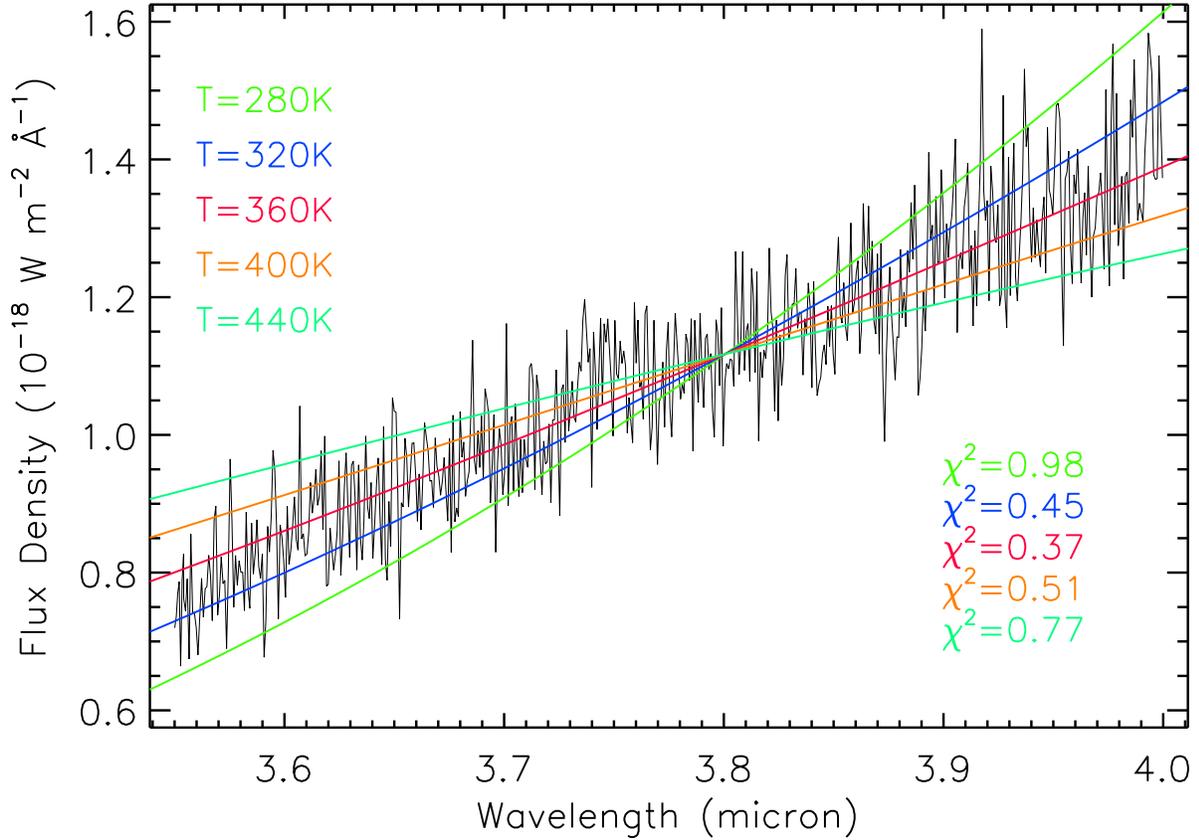}
\caption{ The black line represents the thermal component that obtained after subtraction of a scattering model of the spectrum of 17P/Holmes, which was taken on UT 2007 Oct.\ 27. The colored lines are spectra of blackbodies with different temperatures. The wavelength range (2.8 $<$ $\lambda$ $<$ 3.55$\mu$m) contains a strong water ice absorption band and some emission features, so we only modeled wavelengths $>$ 3.55 $\mu$m. The best fit model, shown in red,  indicates that the color temperature of P/Holmes is 360K.}
\label{plotfour}
\end{center}
\end{figure}

We estimated the slope of the reflectance spectrum of 17P using regions judged to be free from major atmospheric absorptions and potential water ice absorption bands, namely from 0.9 - 1.1$\mu$m, 1.7 - 1.8$\mu$m and 2.2 - 2.3$\mu$m. We made the solar analog spectrum more blue and scaled it to represent the scattering flux. Our scattering model, shown in Figure~\ref{plotthree}, fits the observed spectrum well. Assuming the slope of the comet's albedo with wavelength remains the same from 0.8$\mu$m to 4.2$\mu$m, we extrapolated the fitted spectral slope from the short wavelengths into the thermal wavelengths. Finally, we subtracted the modeled scattered light from the observed spectrum leaving the residual shown in Figure~\ref{plotfour}, which is assumed to be pure thermal emission from the coma. Only the data from UT Oct.\ 27 were used for the thermal modeling, owing to substantial fading of the coma on later dates. 

\begin{figure}[h!]
 \begin{center}
\includegraphics[width=4.5in,angle=90]{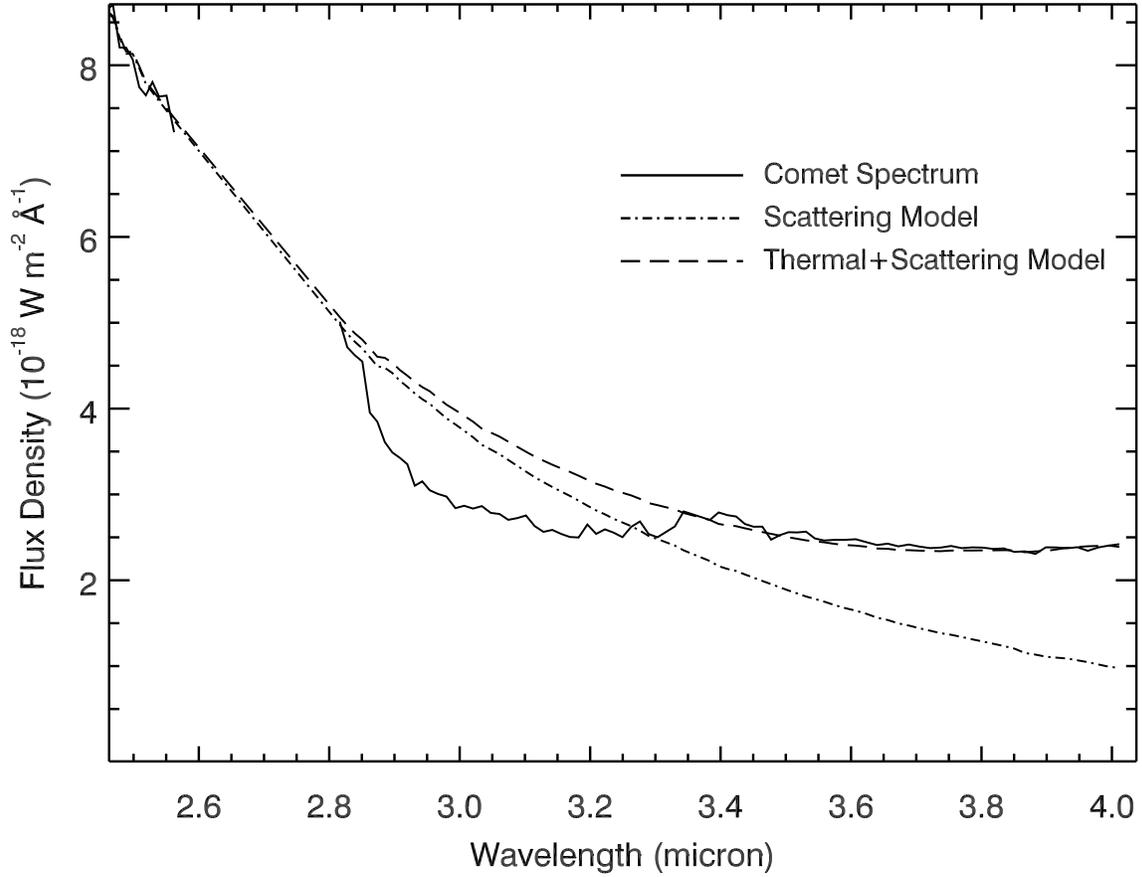}
\vspace{0.1in}
\caption{The solid line shows the comet spectrum, taken on UT 2007 Oct.\ 27.  We do not show the comet spectrum in the wavelength region 2.6$\mu$m - 2.8$\mu$m since it was heavily contaminated by the telluric absorptions. The dash-dot line is the modeled scattering flux. The long dashed line shows the scattering+thermal model that fits the data adequately well. Here the thermal model was calculated using the temperature of 360K.}
\label{plotfive}
\end{center}
\end{figure}

We modeled the thermal flux density, following the method described in \citet{Jewitt:1988}: 

\begin{equation}
f_{BB}(\lambda) = \frac{\epsilon(\lambda)B_\lambda(T_e)C_e}{\Delta^2}
\end{equation}

\noindent where $C_e$ is the emitting  cross-section and $B_{\lambda}(T_c)$ is the Planck function evaluated at the effective grain temperature, $T_e$. The emissivity, $\epsilon(\lambda)$, depends on the ratio of the grain size to the radiation wavelength and other factors that are poorly constrained at present \citep{bohren:1983, mann:2006}. Lacking prior knowledge of the grain sizes, we adopt $\epsilon(\lambda)$ = 1.0 \citep{Gehrz:2005}, which means that the effective grain temperature is equal to the color temperature. The best fitting model, shown in Figure~\ref{plotfour} by a red line, suggests that the color temperature of the thermal component is $T_c$ = 360 $\pm$ 40 K. As shown in Figure ~\ref{plotfive}, our ``scattering+thermal'' model fits the observed comet spectrum adequately well. The considerable uncertainty on T$_c$ results from the fact that only a limited range of thermal wavelengths could be obtained, and that this range does not encompass the Planck maximum.   

For comparison, the expected isothermal blackbody temperature is $T_{BB}$ =180 K at $r$ = 2.4 AU. The ``superheat factor'', $S$, is defined in \citet{Gehrz:1992} to quantify  the ratio between the color temperature of a comet and the temperature of a spherical, co-located isothermal blackbody, which is expressed as \begin{math}S = T_c /T_{BB} \end{math}. Therefore, we find $S$ = 2.0 $\pm$ 0.2 for P/Holmes. Most comets have $S$ in the range of 1 $\leqslant S \leqslant$ 1.5 \citep{Sitko:2004}, but comet Hale-Bopp had $S \sim$ 1.8 \citep{Mason:2001}, similar to the value measured here. 

The superheat depends on many unknown grain properties and it is not possible to use it to make definite statements about the grains. However, it is informative to consider the simplest model, in which the emissivity scales linearly with particle size \begin{math} \epsilon(\lambda) \sim \frac{2\pi a}{\lambda} \end{math} as expected for particles in the Rayleigh limit \citep{bohren:1983}.
 Then, the temperature is related to the emissivity through the energy balance equation for an isothermal sphere:
 
\begin{equation}
\frac{F_\odot}{r^2}(1-A_\lambda) = 4 \epsilon_\lambda \sigma T^4
\end{equation}

\noindent where $F_\odot$ is the solar constant (1.36 $\times$ 10$^3$ W m$^{-2}$), $r$ (AU) is the heliocentric distance, $A_\lambda$ is the effective bond albedo at a wavelength corresponding to the peak of the solar spectrum ($\lambda$ $\sim$ 5000 \AA) and $\epsilon_{\lambda}$ is the effective emissivity at a wavelength corresponding to the peak of the dust emission spectrum (i.e. $\lambda \sim$ 10 $\mu$m for T $\sim $300 K). For a blackbody, A$_{BB}$ = 0, $\epsilon_{BB}$ = 1, we have \begin{math} T_{BB}=(\frac{F_\odot}{4 r^2 \sigma})^{1/4}  \end{math}. And so, the temperature can be written as:

\begin{equation}
T = T_{BB} \left(\frac{1-A_\lambda}{\epsilon_\lambda}\right)^{1/4} 
\end{equation}

\noindent giving 

\begin{equation}
\label{superheat}
S = \left(\frac{1-A_\lambda}{\epsilon_\lambda}\right)^{1/4}.
\end{equation}

\noindent Substituting $S$ = 2, and with $A_\lambda \ll$ 1, Equation (\ref{superheat}) gives $\epsilon_\lambda \sim$ 0.06 and so $a \sim \frac{\epsilon \lambda}{2\pi} \sim$ 0.1 $\mu$m as the characteristic particle size.  This estimate is clearly very crude and the particle size cannot be trusted, probably even to within a factor of a few. However, the important point is that the large superheat suggests that the dominant emitters on UT Oct.\ 27 were small, as they were in Hale-Bopp \citep{Mason:2001}.  
 
\subsection{Water Ice Absorption}
Unlike the depth of an absorption feature, the band profile is relatively insensitive to continuum fitting and thermal emission removal. Therefore, the shape of the absorption features is useful as a constraint on the physical characteristics of the corresponding grains, as discussed in \citet{Sunshine:2007}. The synthetic water ice spectrum was computed based on \citet{Hapke:1981, Hapke:1993}, 

\begin{equation}
\label{ }
R(\mu_0,\mu,\textrm{g})=\frac{w}{4\pi}\frac{\mu_0}{\mu_0 + \mu}\{[1-B(\textrm{g})]P(\textrm{g}) + H(\mu_0)H(\mu)-1\}
\end{equation}

\noindent where R is the bidirectional reflectance, $\mu_0 = cos (i)$ and $\mu=cos (e)$, $i$ and $e$ are the angles of the incident and emergent light, respectively,  $\textrm{g}$ is the phase function between $i$ and $e$, and $w$ is the average single scattering albedo of the surface particles. Parameter $B(\textrm{g})$ is a backscattering function that describes the microstructure of the surface (e.g. porosity, grain size distribution and spacing between particles) \citep{Davies:1997},  $P(\textrm{g})$ is the particle phase function, H($\mu)$ and H($\mu_0)$  are the \citet{Chandrasekhar:1960} H function at angles $\mu$ and $\mu_0$, respectively. To limit the total number of free parameters, all the calculations are at zero phase ($\textrm{g}$ = 0). In our models, we assumed abundant fine grains in the coma, giving B(0) $\sim$ exp(-$w^2/2$), an isotropic phase function [i.e. P(0)=1] and no internal scattering within individual particles. In addition, we assume that the grains in the coma are spherical and smooth. Thus, the internal scattering coefficient S$_I$ is the same as the external scattering coefficient S$_{E}$. We used optical constants of water ice from Warren (1984).

\begin{figure}[h!]
 \begin{center}
\includegraphics[width=4.7in]{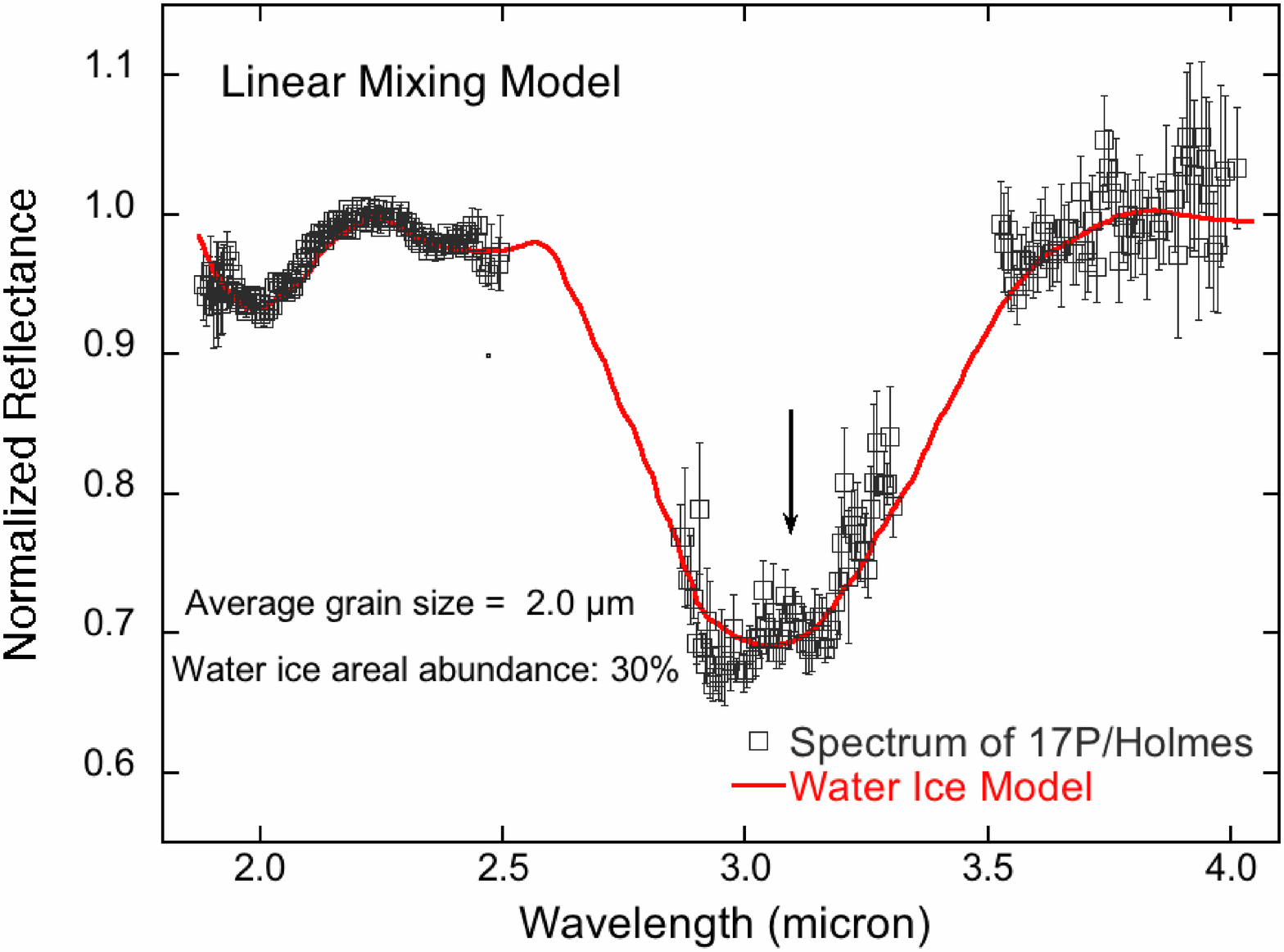} \\
\includegraphics[width=4.7in]{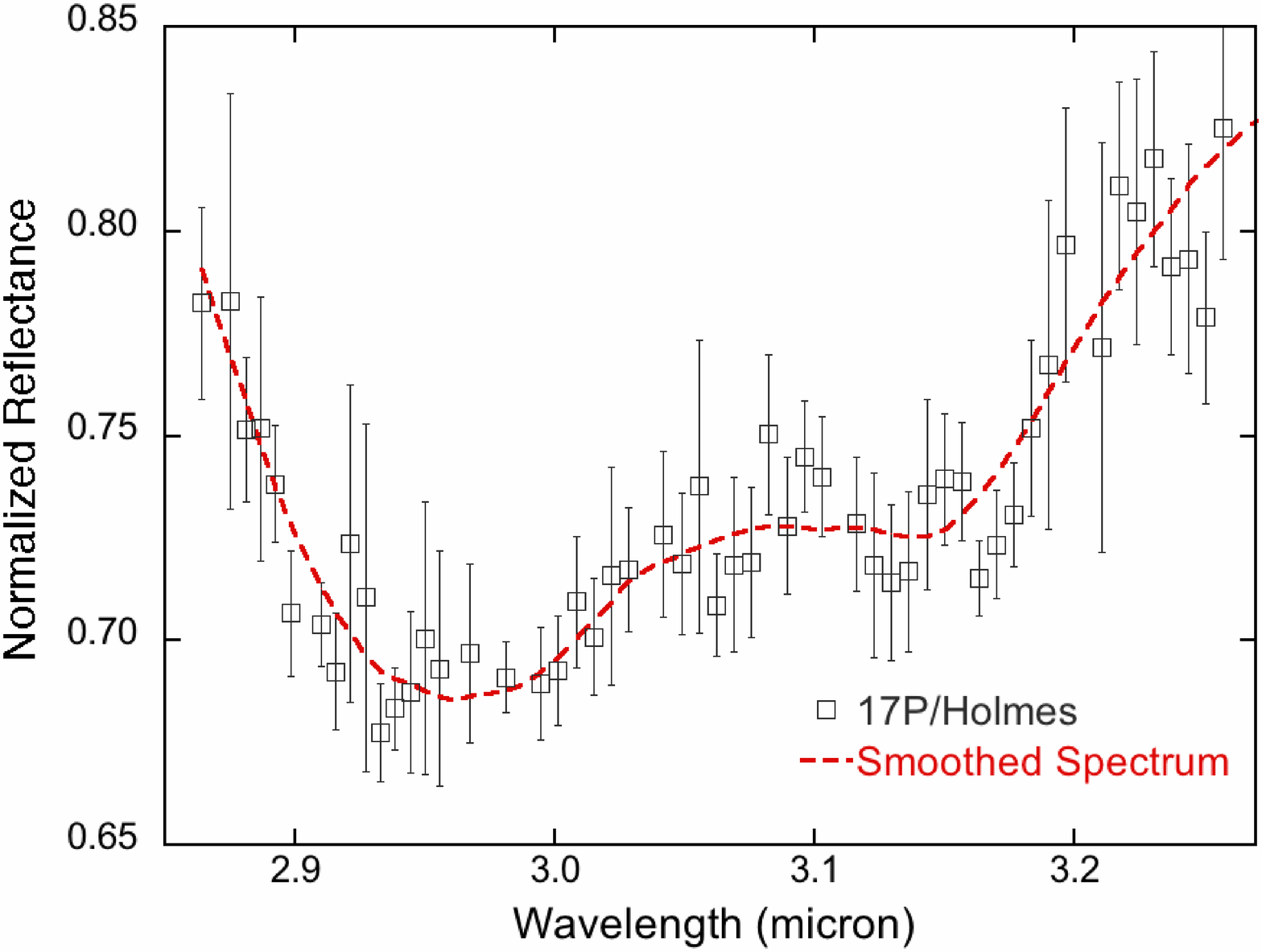}
\end{center}
\caption{a). UPPER  A synthetic water-ice spectrum from the linear-mixing model with 2$\mu$m grains is able to fit the data successfully.  Spectra computed using bigger grains tend to have wider and deeper absorption bands than the ones were observed. The arrow points out a small peak near 3.1$\mu$m, which is due to Fresnel reflection at the surface of ice grains. b). LOWER The black open squares represent the obtained spectrum of the comet on UT 2007 Oct.\ 27.  The red dashed line is a smoothed curve with a smoothing width of 0.1$\mu$m, which clearly shows the Fresnel peak that is about 0.15$\mu$m wide and centered near 3.05$\mu$m. }
\label{plotsix}
\end{figure}

To constrain the abundance and the size distribution of icy grains, we used a ``spatial mixing'' model. The spatial mixing (also known as linear mixing) model, assumed that each component is physically well separated from others and so we linearly mixed the pure water ice with dark and featureless refractory components (e.g. amorphous carbon).  As shown in Figure~\ref{plotsix}a., our water ice model successfully fits the two absorption features, in terms of band centers and overall shapes. In addition, fine structure (the Fresnel peak) was observed near 3.1$\mu$m, as seen in Figure~\ref{plotsix}b, where the red line is a smoothed spectrum to guide the eye. This feature is  due to front-surface reflection from ice grains \citep{brown:2006}, and is observed in some Saturnian icy satellites \citep{clark:2008}. Unfortunately, such a detailed feature is difficult for a single-sized linear mixing model to match. Our best-fit model suggests that the average grain size for 17P/Holmes is about 2$\mu$m and the water ice abundance is about 30\% by surface area. However, one needs to bear in mind that the derived ice abundance is mainly determined from the depth of the 3$\mu$m band, which can be affected by the continuum fitting and the removal of thermal emission. Therefore, our estimate of the water abundance is just a lower limit to the total amount of water ice that could be in the coma. 

The linear mixing model adequately fits the 2 and 3$\mu$m features but it differs from the data at shorter wavelengths ($<$ 1.8$\mu$m). Water ice, the major component in the synthetic model,  produces an additional absorption feature at 1.5$\mu$m, only slightly weaker than the 2.0$\mu$m band. Furthermore, an additional sharp minimum at 1.65$\mu$m, when detected, is widely used as an indicator that the water ice is crystalline in structure.  In the spectra of 17P/Holmes, we detected a fairly strong 2$\mu$m band, however, the 1.5$\mu$m band and the 1.65$\mu$m feature were absent.  

\begin{figure}[h!]
 \begin{center}
\includegraphics[width=5in]{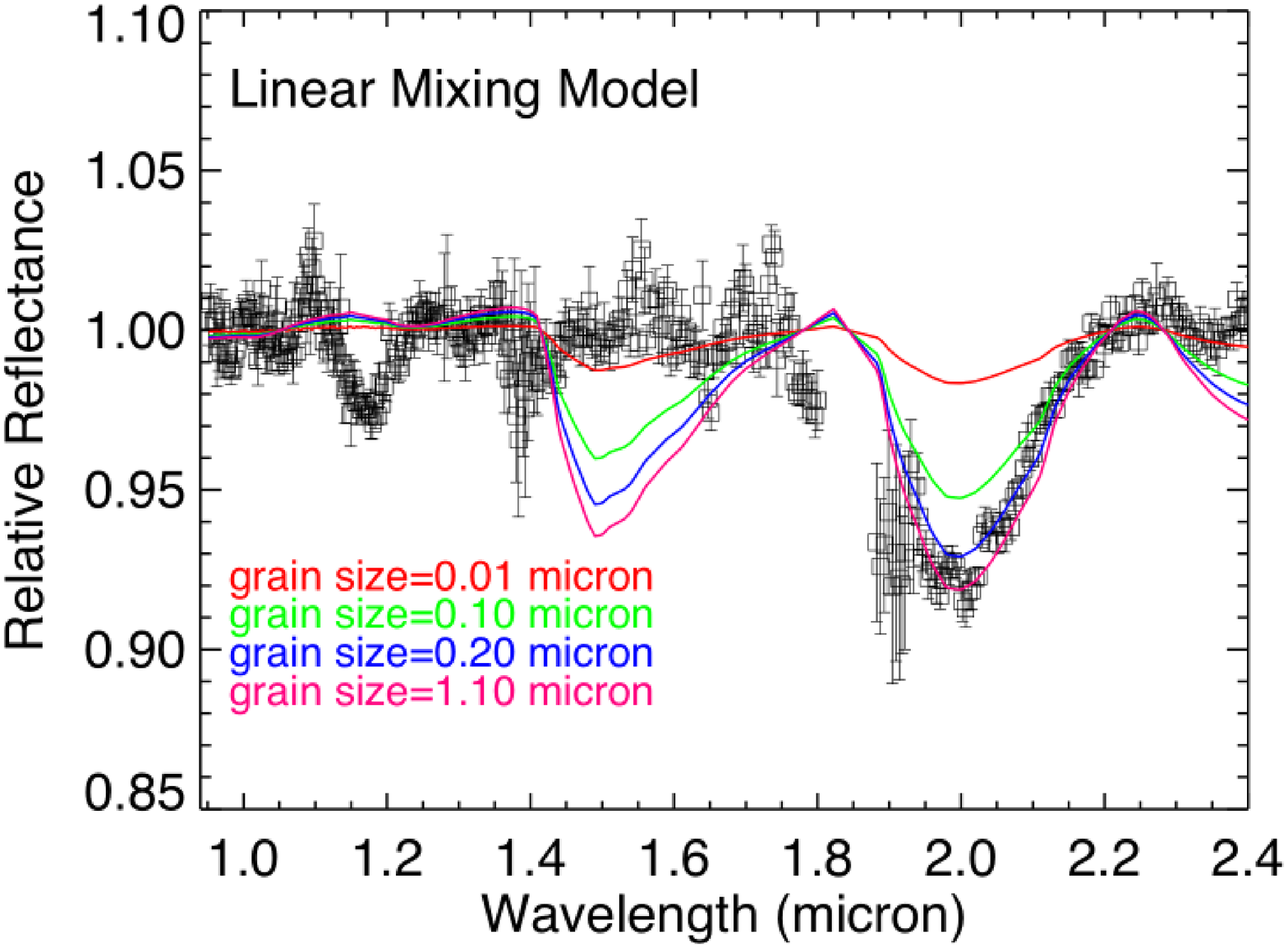}\\
\includegraphics[width=5in]{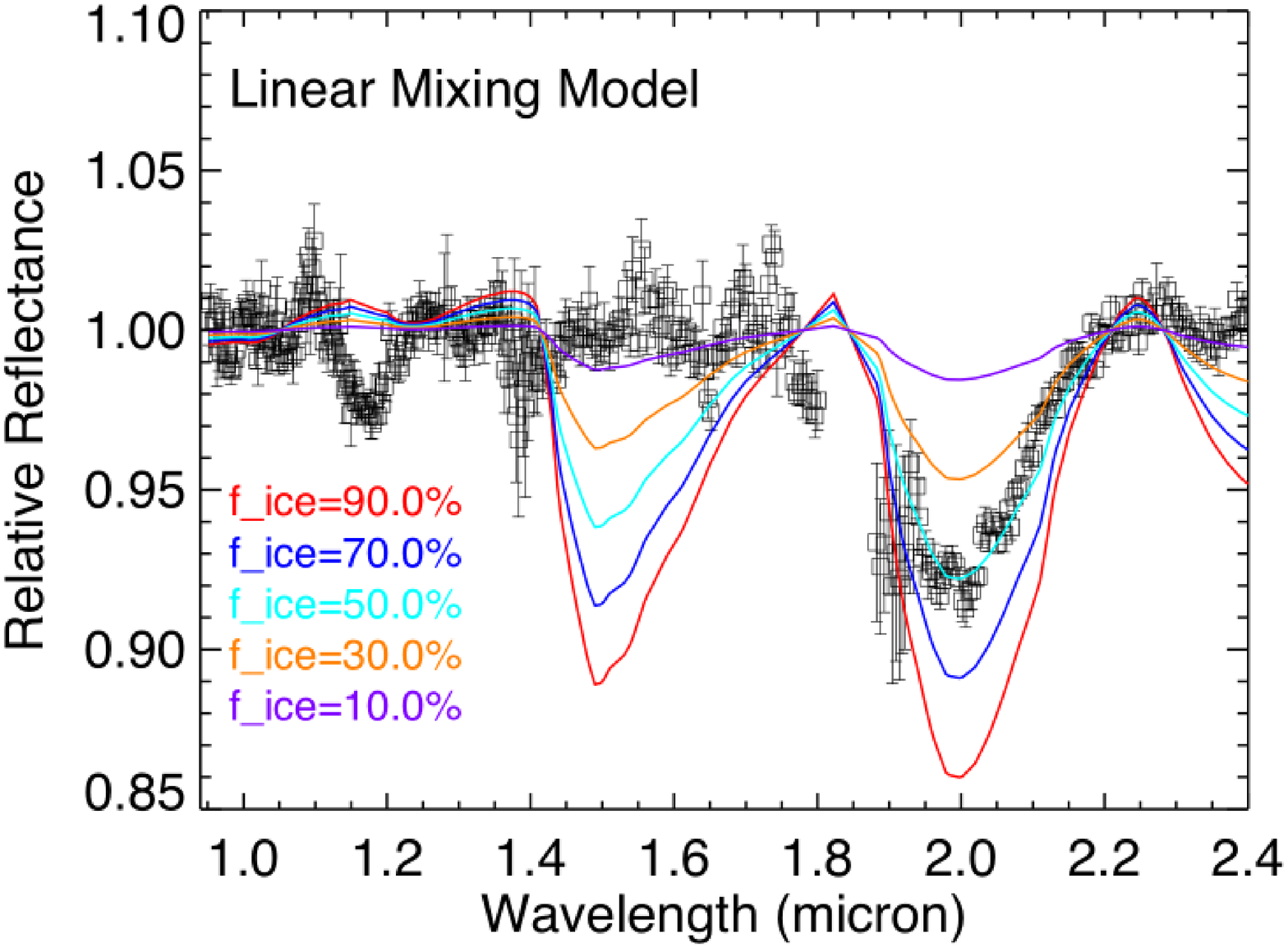}
\end{center}
\caption{ The black line shows the spectrum of the comet on UT 2007 Oct.\ 27 and the colored lines show linear mixing models. a) UPPER  Effect of grain size. b) LOWER Effect of impurities.}
\label{plotseven}
\end{figure}

We explored two ways to decrease the strength of the 1.5$\mu$m band.  First, reducing the grain size can significantly diminish this feature.  However, the strengths of the 1.5$\mu$m and 2.0$\mu$m bands vary together and we found it impossible in our models to selectively weaken the 1.5$\mu$m band while preserving the one at 2$\mu$m (See Figure~\ref{plotseven}a.).  Second, we explored the influence of absorbing impurities in suppression of the 1.5$\mu$m band.  Again,  these experiments affected the 1.5$\mu$m and 2.0$\mu$m bands similarly and we were unable to match the 1.5$\mu$m region of the spectrum of P/Holmes using dirty ice grains (See Figure~\ref{plotseven}b.). Conceivably, some combination of grain size and absorbing impurity could selectively diminish the 1.5$\mu$m band, but the parameter space is huge and we have been unable to reach a solution. 

Is it possible that the 2.0$\mu$m and 3.0$\mu$m bands in the spectrum of P/Holmes do not reflect the presence of water ice? We examined the spectrum of numerous other materials in order to find a match with the P/Holmes spectrum, without success.  We closely considered the possibility that the bands might indicate the presence of a hydrated mineral, rather than pure water ice, since hydrated minerals are known to show a band near 3$\mu$m.  However, the absorption bands in hydrated minerals are shifted by intermolecular forces to wavelengths significantly different from the band centers observed in water ice and in P/Holmes.  For example, the clay mineral montmorillonite shows a deep band starting at 2.8$\mu$m with a minimum near 2.9$\mu$m \citep{2005Icar..179..259R}, incompatible with the comet data. Furthermore, the spectra of hydrated minerals (such as nontronite and palagonite) usually show a sharp band at 1.9$\mu$m  \citep{2007Icar..189..574M}, inconsistent with the 2$\mu$m band in P/Holmes. These different band centers and shapes would be easily resolved in our data. The  positions, shapes and relative depths of the 2.0$\mu$m and 3.0$\mu$m bands point strongly to their origin in water ice rather than in any known hydrated mineral. This conclusion is consistent with in-situ measurements of the dust in comets P/Wild 2 \citep{2006Sci...314.1728K} and P/Tempel 1 \citep{2007Icar..191..223L}, where hydrated minerals are at best a trace species with a very minor impact on the reflected spectrum.  Therefore, we proceed on the interpretation that the 2.0$\mu$m and 3.0$\mu$m bands are indicative of water ice grains, while acknowledging that we are left with the unsolved mystery of why the 1.5$\mu$m band is undetected in P/Holmes.

Although the comet was faint for the LXD observation, we were able to obtain SXD data of high quality on UT Oct.\ 28 and 31, 2007. We observed the 2.0$\mu$m water ice absorption feature in all the SXD data, shown in Figure \ref{plotnine}. The relatively long survival time of this ice feature at the heliocentric distance of 2.4AU provides an extra constraint on the physical properties of the coma grains. To estimate the lifetime of cometary grains, we calculated the sublimation rate as a function of grain size by assuming an equilibrium state between heating (due to absorption of solar radiation) and cooling (due to re-radiation and sublimation) at the surface of the grain: $E_{\odot} =E_{rad} + E_{sub}$.  Here, $E_{\odot}$ (W m$^{-2}$) is the heating rate per unit area, given by: 

\begin{equation} 
\label{sun}
E_{\odot} =\int_{0}^{\infty}\frac{f_{\odot}(\lambda)}{4r^2} Q_{abs}\ d\lambda 
\end{equation}

\noindent where $f_{\odot}(\lambda)$ (W $m^{-2}$ \AA$^{-1}$) describes the solar flux density, r(AU) is the heliocentric distance and Q$_{abs}$ is the absorption efficiency, which was computed using Mie theory.  The quantities $E_{rad}$ (W m$^{-2}$) and $E_{sub}$ (W m$^{-2}$) are the re-radiation rate per unit area and the sublimation rate per unit area respectively, which can be expressed as: 

 \begin{equation}
 \label{rad}
 E_{rad} =\int_{0}^{\infty}B(\lambda, T_s) Q_{e}\ d\lambda  
\end{equation}

\noindent  and 

 \begin{equation}
 \label{sub}
 E_{sub} =HP(T_s)\sqrt{\frac{m}{2\pi kT_s}}
\end{equation}

\noindent where $H$ (J kg$^{-1}$) is the latent heat of sublimation, $P(T_s)$ (N m$^{-2}$) is the equilibrium vapor pressure \citep{1984Icar...60..476F} at the surface temperature $T_s$, m (kg) is the mass of  a water molecule and $k$ (J K$^{-1}$) is the Boltzmann constant. Also, $B(\lambda, T_s)$ is the blackbody radiation at the surface temperature $T_s$ and 
$Q_{e}$  the emission efficiency. Under the assumption of thermodynamic equilibrium between the particle and the surrounding radiation field, we have $Q_{abs} = Q_{e}$ \citep{bohren:1983}.  Then, equations (\ref{sun})-(\ref{sub}) can be compiled in the form:

\begin{equation}
HP(T_s)\sqrt{\frac{m}{2\pi kT_s}}=\int_{0}^{\infty}\left[\frac{f_{\odot}(\lambda)}{4r^2} - B(\lambda, T_s)\right]Q_{abs}\ d\lambda
\end{equation}
 
\noindent We solve this equation to find the grain temperature T$_{s}$ and then compute the specific sublimation rate $\psi$ (kg m$^{-2}$ s$^{-1}$), where $\psi$ = $P(T_s)\sqrt{\frac{m}{2\pi kT_s}}$.  Consequently, we can calculate the lifetime of an ice grain by \begin{math}\tau=\rho \int_{a_{m}}^{a_M}\psi^{-1} da \end{math}, where $\rho$ is the water ice density and $a_{m}$, $a_{M}$ are the minimum and maximum grain diameter, respectively.  

\begin{figure}[h!]
\begin{center}
\includegraphics[width=4.5in,angle=90]{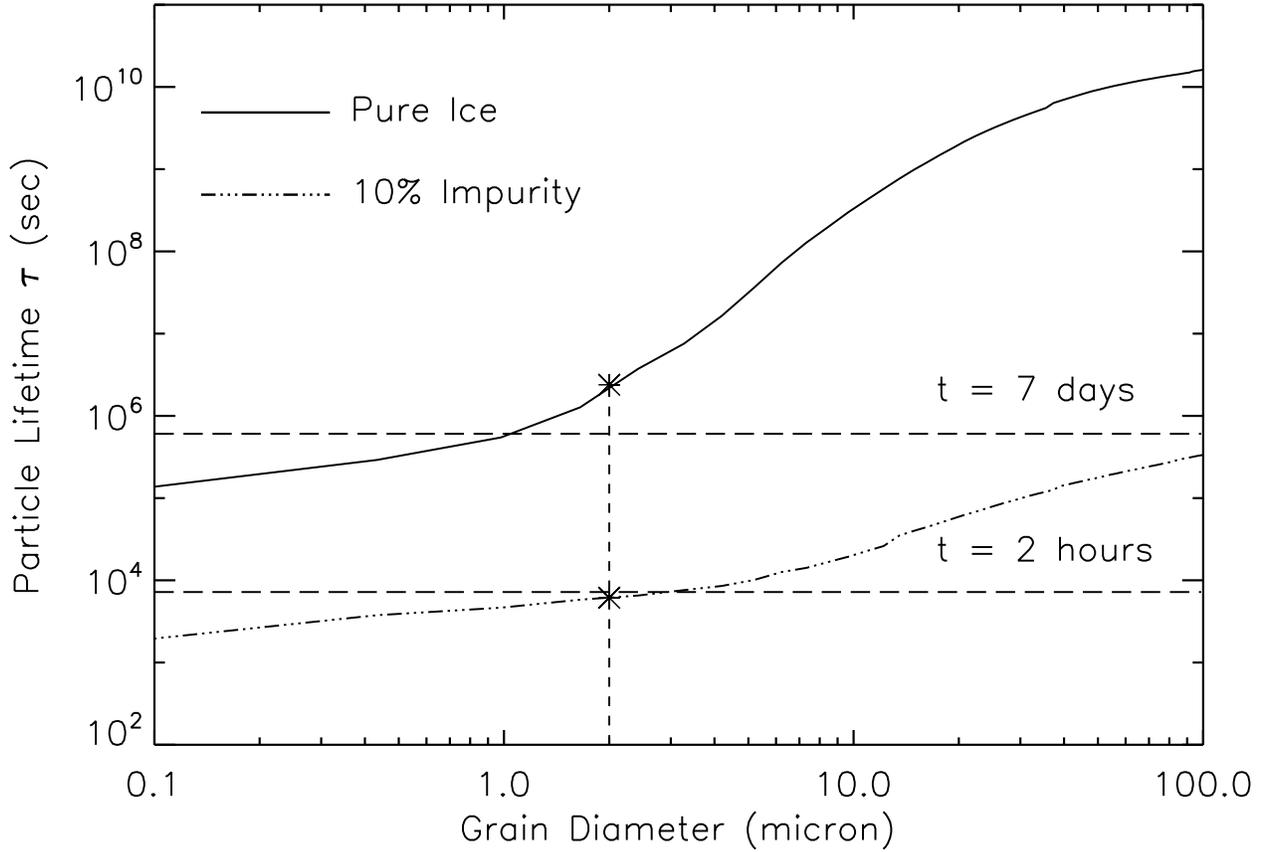}
\end{center}
\caption{ Lifetimes of pure and dirty ice grains at the heliocentric distance of 2.45 AU. The two ``star'' symbols show the lifetimes of dirty (lower) and clean (upper) ice grains with a diameter of  2.0$\mu$m. The lower horizontal dashed line indicates the required in-slit residence time which sets the lower limit of the lifetime for ice grains in the coma of 17P. It shows that only pure water ice was able to survive and to produce the detected absorption features. The upper horizontal dashed line represents the time interval between the start of the outburst and the end of our observations.}
\label{ploteight}
\end{figure}

For pure ice grains, the sublimation rate was calculated using optical constants from Warren (1984). However, the observed cometary grains may not be pure but contaminated by refractory particles, such as amorphous carbon. Impurities can significantly shorten the lifetime of an icy grain by reducing the albedo and increasing the equilibrium temperature in sunlight.  The optical constants of amorphous carbon from \citet{1991JRASC..85..201R} were used as the darkening material in the dirty ice and the Maxwell-Garnett mixing rule \citep{1986A&A...164..397M} was applied.  The sublimation lifetimes of pure and dirty water ice grains as functions of their size are shown in Figure~\ref{ploteight}.

First of all, the presence and strength of the water ice absorption features indicates that the fresh icy grains survived against sublimation over a substantial fraction of the projected area of the slit. We extracted spectra of pure coma from a region that is 1.$\!\!^{\prime\prime}$5 to 3$''$ away from the nucleus.  Although the SNR of the outer coma became much weaker, we found the water ice features persisted in the coma spectra. Our result suggests that the ice grains survived the sublimation up to 3$''$ away from the nucleus. The geocentric distance was 1.62 AU at the time of the observation, so that the projected distance of 3$''$ on the coma corresponds to 3534 km. The measured expansion speed of the coma is about 500 m/sec (Rachel Stevenson  2009, private communication), and so it would take about 2 hours for newly released icy grains to travel from the nucleus to the region that is 3.5 thousand kilometers away, if moving parallel to the length of the slit. As noted above, the best-fit to the band shapes at 2$\mu$m and 3$\mu$m gives an average diameter of coma ice particles of $\sim$ 2$\mu$m. Figure~\ref{ploteight} shows that dirty ice grains of this size are not able to survive for 2 hours, the residence time in the slit. More importantly, the source of the observed ice grains likely survived for at least  7 days after the initial outburst. The observed long-lasting water ice absorption feature suggests that the ice grains in the coma of 17P must be relatively clean. 
 
\subsection{Spectral Slope}
\begin{figure}[h!]
 \begin{center}
\includegraphics[width=4.5in, angle=90]{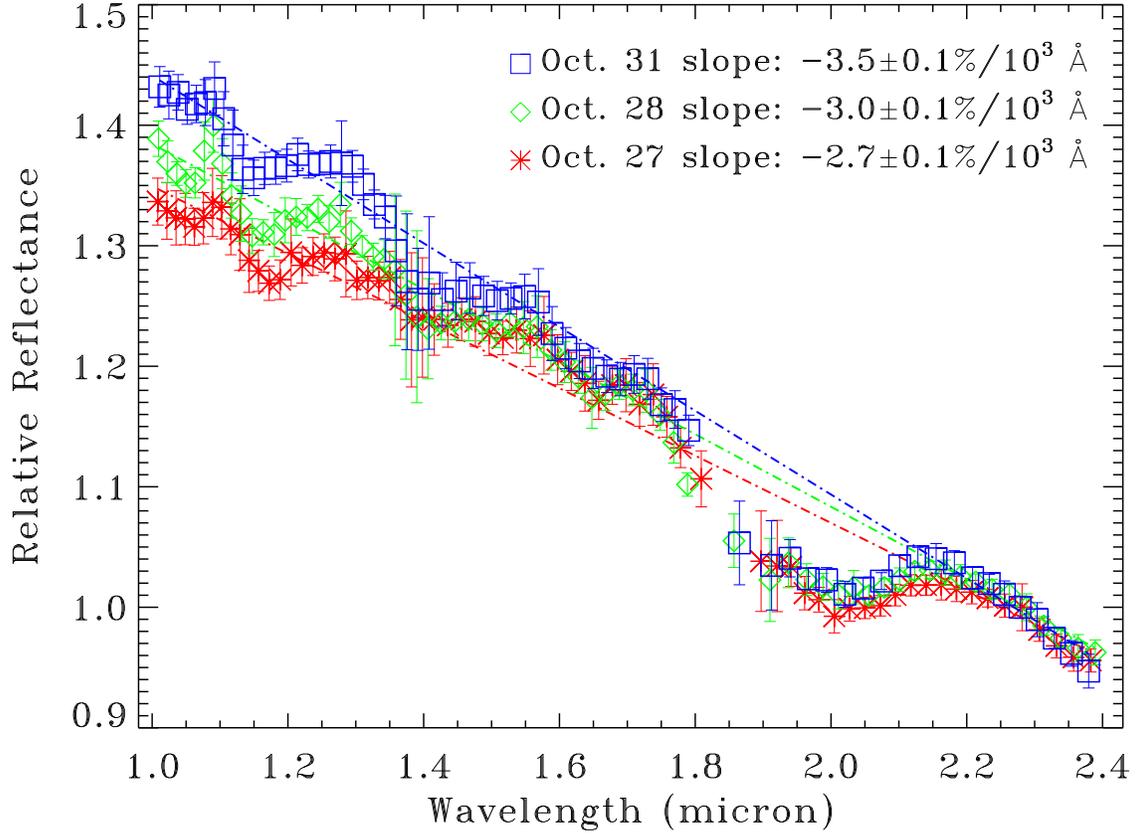}
\end{center}
\caption{Red, green and blue points represent the SXD spectra of 17P/Holmes taken on UT 2007, Oct.\ 27, 28 and 31,  which are normalized to unity at 2.2$\mu$m. Dashed lines are linear spectral slopes, which reflect either the chemical composition or the size distribution of the cometary coma. The figure illustrates the spectral slope becomes steeper with time further away from the original outburst.  }
\label{plotnine}
\end{figure}

The negative slope (Figure ~\ref{plotnine}) distinguishes 17P from most comets that generally show neutral or positive slopes at optical and near infrared wavelengths \citep{Jewitt:1986}. Previous studies have shown that near infrared spectra of pure water ice or hydrated minerals also have negatively sloped continua within the same wavelength range (Clark et al. 1986). The spectral slopes in P/Holmes grew steeper with time from the initial outburst, as shown in  Figure~\ref{plotnine}, suggesting that either the abundance of icy grains was increasing or the average grain size was decreasing between the nights. Since the observations were made after the outburst and the gas outflows became weaker with time, the shrinkage of grains is more likely to be responsible for the change of the spectral slope. The scattering efficiency, $Q_s$, generally depends on the ratio of the particle size to the wavelength.  When the grain sizes are comparable to or slightly smaller than the wavelength, the scattering efficiency varies inversely with the wavelength.  Therefore, as the average size of coma particles becomes smaller, the blue scattering efficiency increases and the reflectance spectrum becomes steeper.

\subsection{Other Spectral Features}
\begin{figure}[h!]
\begin{center}
\includegraphics[width=5.0 in]{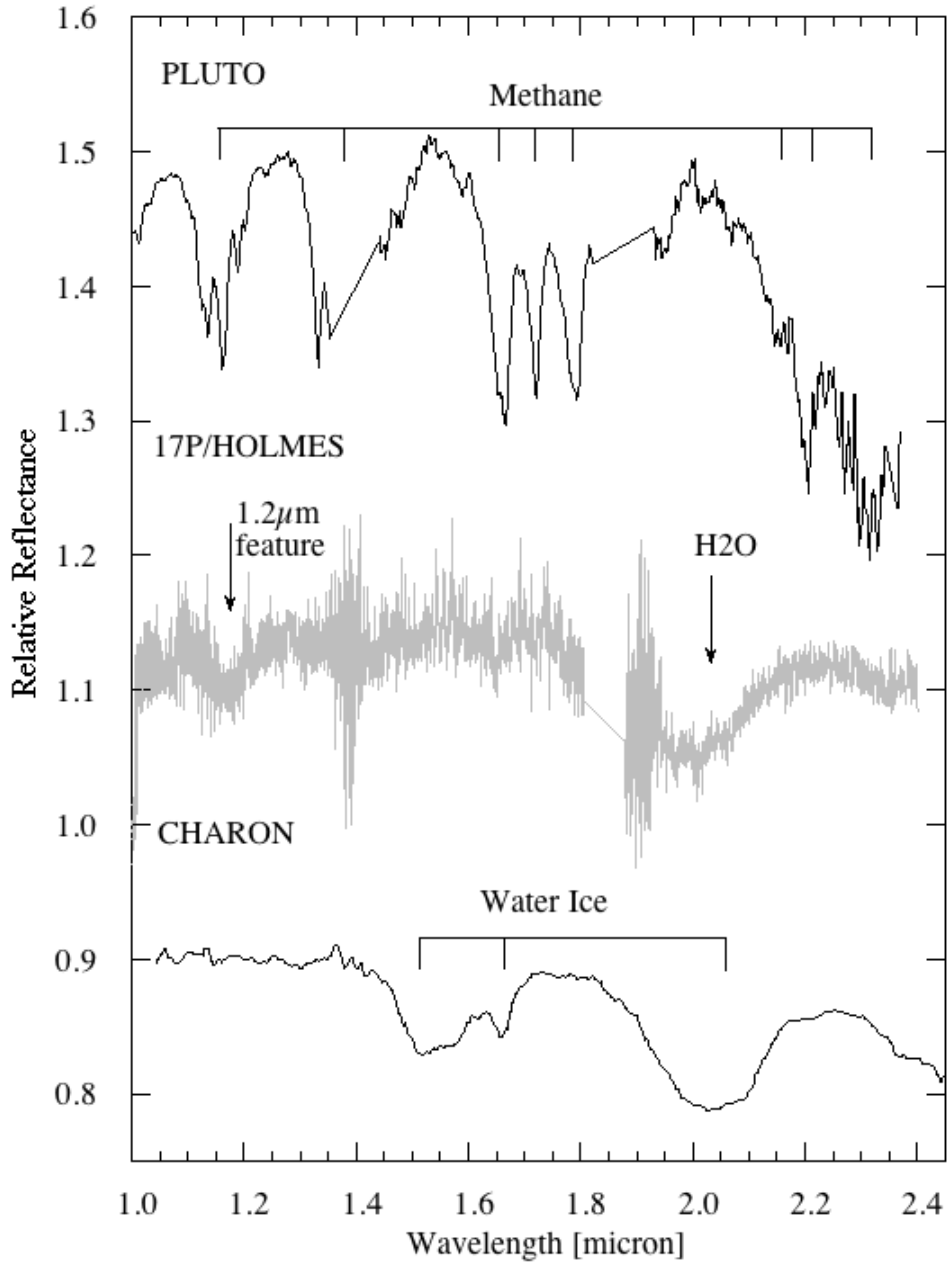}
\end{center}
\caption{Comparison of the NIR spectrum of 17P/Holmes, Pluto and Charon (from \cite{2000Sci...287..107B}). The spectrum of Pluto is dominated by methane and water ice dominates the spectrum of Charon. The significant absorption feature near 1.2$\mu$m in the spectrum of the comet is not associated with either water ice or methane.}
\label{plotten}
\end{figure}

In addition to water ice absorptions, a broad absorption feature was found near 1.2$\mu$m in the spectrum of 17P/Holmes. To verify the presence of  this absorption feature, we examined the SXD data of 17P/Holmes obtained on UT 2007, Oct.\ 27, Oct.\ 28 and Oct.\ 31, using two different solar analog stars and the flux calibrated solar spectrum to calculate the relative reflectance.  We found that the 1.2$\mu$m feature appeared consistently over three nights and its existence was independent to the choice of the solar analog. Furthermore, the center and the overall profile of this absorption feature are also insensitive to the particular solar analogs used. Some alkanes, such as methane and ethane, show an absorption band near 1.2$\mu$m attributed to the second overtone of C-H stretch.  However,  these hydrocarbons also show other, stronger features near 1.7 and 2.3$\mu$m that do not appear in the spectrum of 17P/Holmes (see Figure~\ref{plotten}). So, we do not favor an interpretation in terms of hydrocarbons. Furthermore, we checked for this feature amongst the diffuse interstellar bands (DIBs) \citep{1995ARA&A..33...19H}, but found no matches. At present, the 1.2$\mu$m feature remains unidentified. 

\begin{figure}[h!]
 \begin{center}
\includegraphics[width=5.0in,angle=90]{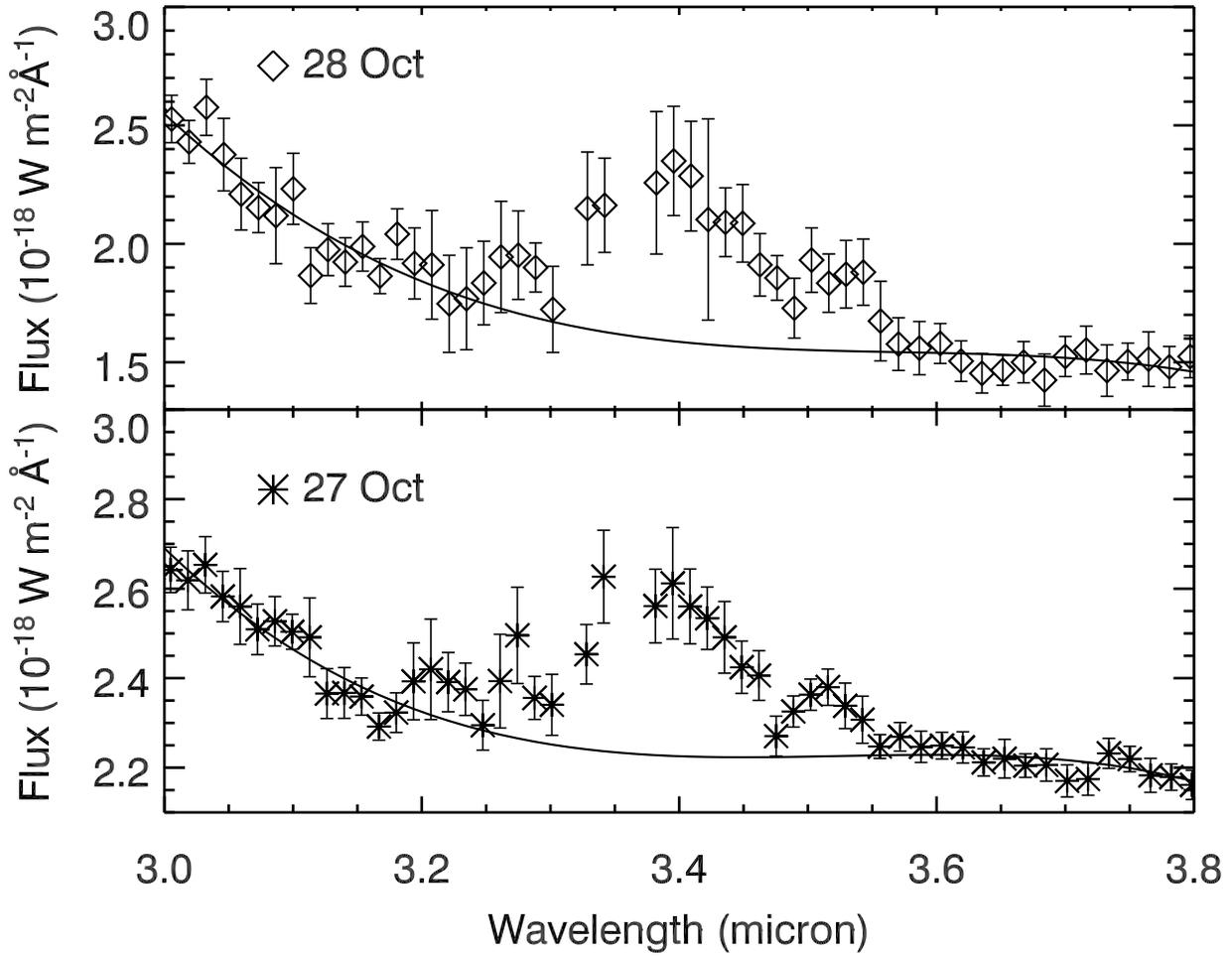}
\end{center}
\caption{ Points are observed spectra of 17P/Holmes and solid curves are best fits to the continuum. On both nights, the emission features were distinct from the continuum, with a strong peak near 3.36$\mu$m and a weak peak at 3.52$\mu$m. }
\label{ploteleven}

\end{figure}

Significant emission features were detected near 3.4$\mu$m in spectra of 17P/Holmes from UT 2007 Oct.\ 27 and 28. Although the data of UT 2007 Oct.\ 28 appeared much more noisy than from UT 2007 Oct.\ 27, the 3.4$\mu$m emission feature was strong enough to be discerned clearly from the noisy continuum.  The LXD data of UT 2007 Oct.\ 31 suffered from very low SNR and thus were not used.  As shown in Figure~\ref{ploteleven}, the peak wavelength of  the stronger emission band is near $\sim$ 3.36$\mu$m and the width is $\sim$ 0.15$\mu$m. In addition, there is a weaker feature centered at 3.52$\mu$m with the width of  0.07$\mu$m.  These characteristics are consistent with the $\nu_2$,  $\nu_3$ and  $\nu_9$ CH-stretching bands of methanol, which were first observed in comet Halley two decades ago and seen in several comets since then \citep{hoban:1991, reuter:1992, mumma:2001}. We calculated the methanol production rate following the method described in \citet{1995Icar..116....1D}. The derived production rate is about 2.2 $\pm$ 0.5 (10$^{27}$mol/s), which is comparable with those of other comets. However, we note that the methanol production rate measured in \citet{2008ApJ...680..793D} is several times higher than the value we obtained in this work. This difference may due to the timing difference between their observations and ours. Moreover, the measurements of Dello Russo et al. are based on high-resolution observations. For these reasons, we do not see a significant contradiction between their results and ours.

\section{Discussion}
Our thermal model suggests that the color temperature of the coma of P/Holmes is about 360K, however, water ice absorption features were also detected in the spectra at the same time. Given that the life time of ice grains decreases as the temperature increases, the presence of  ice grains requires a low temperature (T $<$ 270 K) to ensure the survival of solid water.  One explanation for this temperature dichotomy is that the coma of  P/Holmes might contain two distinctive grain populations that have different temperatures not in thermal equilibrium at the time of the observations. The temperature of a particle at a given heliocentric distance is determined by many factors, including the chemical composition and the grain size.  For sufficiently small absorbing grains (e.g. carbonaceous compounds), the size exerts a dominant influence on the temperature \citep{Hanner:1983}. Most particles made of transparent materials have equilibrium temperatures  below that of a co-located blackbody regardless of their size \citep{Kolokolova:2004}. Compared with carbon, pure water ice has much higher heat capacity and albedo. If water ice grains and refractory grains are not in thermal contact, they could maintain distinct temperatures. Therefore, our observations indicate that the coma of 17P/Holmes consists of distinct cold, icy grains and hot fine refractory grains. 

Interestingly, our inference of the coexistence of cold and hot particles in the coma of 17P/Holmes is consistent with observations of P/Tempel ~1 from NASA's ``Deep Impact'' (DI) mission. There, a strong water ice absorption at 3$\mu$m was observed in the spectra of the ejecta within 3 sec of the impact. Synthetic mixing models revealed small, pure particles with an average diameter of 1$\mu$m \citep{Sunshine:2007}. Simultaneously, a significant thermal component from hot dust was observed along with the ice features in the DI spectra at wavelengths $>$ 3$\mu$m. Therefore, the impact liberated ejecta is likely to contain both pure water ice grains and hot dust particles. In addition, the survival of water ice features throughout the DI flyby observations indicates that the ice grains were thermally/physically separated from other components and therefore pure \citep{Sunshine:2007}. 

In other comets, fragmentation of grains is believed to be an important process. For example, the distribution of impacts of small dust grains onto spacecraft in the comae of P/Halley and P/Wild 2  \citep{1987A&A...187..824B, 2004JGRE..10912S03C} was highly non-random, suggesting spatially correlated swarms of particles that could have been produced by fragmentation of larger parent particles.  Fragmentation could result from electrostatic disruption caused by the build-up of charge on aggregate dust grains exposed to the ionising radiation of the Sun.  Alternatively, fragmentation could be due to the loss of a binding agent (``glue'') holding sub-grains together in the initial aggregate.

These considerations suggest  a simple picture in which the body of 17P/Holmes consists of porous aggregates of refractory particles and ice grains held together by ice and/or organic compounds (the latter suggested by the detection of the 3.4$\mu$m methanol feature; see also \citet{1988Icar...76..100L}). Once ejected from the nucleus into sunlight, the ice begins to sublimate and the aggregates break up, releasing the particles we detected and contributing to the dramatic increase in brightness of the coma. A porous aggregate structure of low tensile strength could be representative of the nucleus as a whole, and fragmentation of ejected pieces of the nucleus might explain the multiple secondary sources observed in high-contrast imaging of the coma.

\section{Summary}
Near infrared spectral observations of comet 17P/Holmes were obtained within a week of its UT 2007 Oct.\ 24 outburst, with the following results:

\begin{itemize}

   \item Two deep absorption features were observed at 2$\mu$m and 3$\mu$m respectively, consistent in central wavelength, band width and shape with the presence in the coma of water ice grains. Spectral fits with linear mixing models show that 17P/Holmes contains a significant fraction of pure water ice and that the central coma is dominated by fine ice grains of micron size.  
   
   \item The spectral fits with water ice are so good that we find it difficult to believe that the 2$\mu$m and 3$\mu$m bands could have any other explanation.  Indeed, an explicit search for hydrated and other minerals that have similar features failed to produce a convincing match.  However, with water ice models, we have yet to explain why the 1.5$\mu$m band of water ice is absent from our data.
 
  \item We observe strong thermal emission at wavelengths $\lambda >$ 3.2$\mu$m,  which we attribute to hot dust. The derived color temperature is 360 $\pm$ 40K and the corresponding superheat factor is $\sim$ 2.0. As in other comets (e.g. C/Hale-Bopp)  the high superheat indicates that the thermal emission is dominated by small, refractory grains. 
    
    \item The simultaneous presence of hot, refractory grains and water ice grains (which must be cold in order to minimize sublimation) shows that these two populations are not in thermal contact.  
    
    \item A solid-state absorption band observed near 1.2$\mu$m on three nights remains unidentified. Further laboratory experiments are needed. 
    
    \item Our observations of 17P/Holmes resemble those of P/Tempel~1 from NASA's Deep Impact mission, in the sense that both comets showed small, pure ice grains 
coexisting with hot refractory grains in the coma. 
     
    \end{itemize}

\section{Acknowlegement}
We thank Alan Tokunaga for scheduling these observations on short notice and John Rayner and Mike Cushing for help with the data processing. Joshua Emery and the referee, Yan Fern\'andez, offered constructive remarks and we thank Zahed Wahhaj and Pedro Lacerda for reading and commenting upon the manuscript. This work was supported by grant NNG06GG08G to David Jewitt from the NASA Planetary Origins program.

\newpage

\newpage

\begin{deluxetable}{lccclclc}\tablewidth{5.8in}
\tabletypesize{\scriptsize}
\tablecaption{ Observational parameters 
  \label{obstable}}
\tablecolumns{8} \tablehead{\colhead{ UT Date }  &
\colhead{Wavelength}    & \colhead{Itime} & \colhead{Airmass} &
\colhead{$r$}&\colhead{$\Delta$}& \colhead{$\alpha$}&CSO Tau \\
\colhead{}&\colhead{$\mu$m}&\colhead{(s)}&\colhead{}&\colhead{AU}&\colhead{AU}&\colhead{deg}&\colhead{}}
\startdata
2007 Oct. 27 &  0.8-4.2 & 1440 (SXD) +  780 (LXD) & 1.16 - 1.32 & 2.44 & 1.63 & 16.2 & 0.05 \\
2007 Oct. 28 &  0.8-4.2 & 1200 (SXD) +  900 (LXD) & 1.29 - 2.30 & 2.45 & 1.62 & 15.9 & 0.04\\
2007 Oct. 31 &  0.8-4.2 &   720 (SXD) +  450 (LXD) & 1.16 - 1.19 & 2.46 & 1.62 &  15.1&0.06\\
\enddata

\end{deluxetable}

\clearpage

\clearpage

\clearpage

\clearpage

\clearpage

\clearpage

\clearpage

\vspace{2cm}

\end{document}